\documentclass{ws-procs9x6}

\begin{document}

\title{Dynamical study of the pentaquark antidecuplet in a constituent
quark model
\footnote{\uppercase{T}alk  given at the 
\uppercase{I}nternational \uppercase{W}orkshop \uppercase{PENTAQUARK04}, 
 \uppercase{SP}ring-8, \uppercase{H}yogo, 
\uppercase{J}apan, \uppercase{J}uly 20-23, 2004}}

\author{Fl. STANCU
}

\address{ University of Li\`ege,\\
Institute of Physics, B.5, Sart Tilman,\\ 
B-4000 Li\`ege 1, Belgium\\ 
E-mail: fstancu@ulg.ac.be}

\maketitle

\abstracts{Dynamical calculations are performed for all members of
the flavor antidecuplet to which the pentaquark
$\Theta^+$ belongs. The framework is a constituent quark model
where the short-range interaction has a
flavor-spin structure. 
From symmetry considerations the lowest state acquires a positive parity.
By fitting the mass of  $\Theta^+$ of minimal content
$uudd\overline{s}$,
the mass of $\Xi^{--}$, of minimal content $ddss \overline{u}$,
is predicted to be approximately 1960 MeV. It is shown that the octet and
antidecuplet states with the same quantum numbers mix ideally 
due to SU(3)$_F$ breaking.}

\section{Introduction}

At present there is a large variety of approaches to pentaquarks:
chiral soliton or Skyrme models, constituent quark models, instanton 
models, QCD sum rules, lattice calculations, etc. Here I shall discuss
the pentaquarks in the framework of constituent quark models. 
These models describe a large number of observables 
in ordinary hadron spectroscopy as e. g. spectra, static properties,
decays, form factors, etc. Therefore it seems interesting to look for their
predictions for exotics. 
I shall refer to two standard constituent quark models: the color-spin (CS) 
model where the hyperfine interaction is of one-gluon exchange type and
the flavor-spin (FS) where the hyperfine interaction is due to
meson exchange. There are also hybrid models where the hyperfine 
interaction is a superposition of CS and FS interactions.

Presently the main issues of any approach to pentaquarks are: 
\begin{enumerate}
\item
The spin and parity
\item
The mass of $\Theta^+$  
(or $\Theta^0_c$ for heavy pentaquarks)
\item
The splitting between isomultiplets of a given SU(3)$_F$ representation
\item
The mixing of representations due to SU(3)$_F$ breaking
\item
The width
\item
The production mechanism
\end{enumerate}
\noindent
Here I shall present results for light and heavy pentaquarks
obtained in the FS model. I shall cover all but the last two items.
\section{Parity and spin }
 
The antidecuplet to which $\Theta^+$ belongs \cite{DPP} can be obtained
from the direct product of two flavor octets, one representing a baryon ($q^3$)
and the other a meson ($q {\overline q}$) 
\begin{equation}
8_F \times 8_F = 27_F + 10_F +{\overline {10}}_F + 2(8)_F + 1_F 
\end{equation}
The antidecuplet ${\overline {10}}_F$  can mix with  $8_F$, for example, because
SU(3)$_F$ is not exact.  This  mixing will be considered below.

To find the parity of $\Theta^+$ and of its partners one 
looks first at the $q^4$ subsystem with $I$ = 0 and $S$ = 0, i. e. with quantum
numbers compatible with the content $uudd {\bar s}$ of  $\Theta^+$. This means
that the flavor and spin wave functions have symmetry $[22]_F$ and 
$[22]_S$ respectively.
Their direct product can generate the state $[4]_{FS}$. If
the orbital wave function contains a unit of orbital excitation, it is
described by $[31]_O$. The color part of $q^4$ is
$[211]_C$ in order to give rise to a color singlet state after the coupling 
to ${\bar q}$.
Then  $[31]_O \times [211]_C \rightarrow [1111]_{OC}$ so that 
the Pauli principle requires $[4]_{FS}$. In the FS 
model the contribution of the hyperfine attraction of $[4]_{FS}$ is so strong that it
fully overcomes the excess of kinetic energy due to the orbital excitation
and the lowest state has positive parity irrespective of the flavor content
of the pentaquark \cite{FS1}. The spin is either 1/2 or 3/2.
  
In the CS model a positive parity as lowest state is also possible
in principle \cite{JM}. The orbital and flavor symmetries give 
$[31]_O \times [22]_F \rightarrow [211]_{OF}$ and, due to Pauli
principle, this must combine with $[31]_{CS}$, the lowest allowed symmetry.  
Then only if the excess of kinetic energy is compensated 
by the attractive hyperfine interaction  the parity would be positive.
This does occur in realistic calculations with CS interaction,
as implied by Refs. \cite{ENYO,TS}, so that the lowest state
has negative parity. 
More generally,
parity remains a controversial issue also in QCD
lattice calculations \cite{SASAKI,CH1}.  
QCD sum rules lead to negative parity \cite{SUMRULE}.

\section{The orbital wave function}
There are four internal Jacobi coordinates
$\vec{x} = {\vec{r}}_{1} - {\vec{r}}_{2}$,
$ \vec{y}\ =\
{\left({{\vec{r}}_{1}\ +\ {\vec{r}}_{2}\ -\ 2{\vec{r}}_{3}}\right)/\sqrt
{3}}$,
~$\vec{z}\ =\ {\left({{\vec{r}}_{1}\ +\ {\vec{r}}_{2}\ +\ {\vec{r}}_{3}\ -\
3{\vec{r}}_{4}}\right)/\sqrt {6}}\ , \,$ and 
$\vec{t}\ =\
{\left({{\vec{r}}_{1}\
+\ {\vec{r}}_{2}\ +\ {\vec{r}}_{3}+\ {\vec{r}}_{4}-\
4{\vec{r}}_{5}}\right)/\sqrt {10}}$~where 5 denotes the antiquark.
The total wave function is a linear combination of three independent 
orbital basis vectors contributing with equal weight\cite{FS1,FS2}
~${\psi }_{1}\ = \ \psi_0\ \ z\ {Y}_{10}\ \left({\hat{z}}\right)$,
${\psi }_{2}\ = \ \psi_0\ \ y\ {Y}_{10}\ \left({\hat{y}}\right)$ and
${\psi }_{3}\ = \ \psi_0\ \ x\ {Y}_{10}\ \left({\hat{x}}\right)$ where
\begin{equation}\label{PSI0}
\psi_0 = {[\frac{1}{48 \pi^5 \alpha \beta^3}]}^{1/2}
\exp\ \left[{-\ {\frac{1}{4 \alpha^2}}\ \left({{x}^{2}\ +\
{y}^{2}\ +\ {z}^{2}}\right)\ -\ {\frac{1}{4 \beta^2}}\ {t}^{2}}\right]~.
\end{equation} 
contains two variational parameters:~ $\alpha$, the same for all $q^4$ 
coordinates $\vec{x}$, $\vec{y}$ and $\vec{z}$,
and  $\beta$, related to  $\vec{t}$, the relative coordinate
of $q^4$  to $\overline{q}$.
\section{The antidecuplet}
The pentaquark masses are calculated by using the realistic Hamiltonian
of Ref. \cite{GPP}, which leads to a good description of low-energy
non-strange and strange baryon spectra. It contains an internal kinetic
energy term $T$, a linear confinement potential $V_c$ and a short-range
flavor-spin hyperfine interaction $V_{\chi}$ with an explicit radial form 
for the pseudoscalar meson exchange. Details of 
these calculations are given in Ref. \cite{FS2}.
\begin{table}
\tbl{The hyperfine interaction $ V_{\chi}$  
integrated in the flavor-spin space. 
for some $q^4$ subsystems (for notation see text). }
{\footnotesize
\begin{tabular}{@{}c|c|c@{}}
\hline
{} &{} &{} \\ [-0.5ex]
$\bf {q^4}$ & $\bf {I,~ I_3}$ &  $\bf {V_{\chi}}  $ \\[1ex]
\hline
$\bf {uudd}$ & $\bf {0,~ 0}$    & $\bf {30~ V_{\pi} - 2~ V^{uu}_{\eta} - 4~ V^{uu}_{\eta'}} $\\[1ex]
$\bf {uuds}$ & $\bf {1/2,~1/2}$ & $\bf {15 V_{\pi} - V^{uu}_{\eta} - 2~ V^{uu}_{\eta'}
+ 12~ V_K + 2~ V^{us}_{\eta}  - 2~ V^{us}_{\eta'}}$ \\[1ex]
$\bf {ddss}$ & $\bf {1, -1}$ & $\bf {V_{\pi} + \frac{1}{3} V^{uu}_{\eta} + \frac{2}{3}V^{uu}_{\eta'} 
+ \frac{4}{3}  V^{ss}_{\eta} + \frac{2}{3} V^{ss}_{\eta'} + 20 V_K}$ \\[1ex]
 &  & + $\bf {\frac{16}{3} V^{us}_{\eta} - \frac{16}{3} V^{us}_{\eta'}}$ \\[1ex]
\hline
\end{tabular}\label{FOURQ} }
\end{table}
The expectation values of the hyperfine interaction $ V_{\chi}$ integrated 
in the flavor-spin space,
are shown in Table \ref{FOURQ}
for the three $q^4$ subsystems necessary to construct the antidecuplet. They 
are expressed in terms of the two-body radial form 
$V^{q_a q_b}_{\gamma}$ of Ref. \cite{GPP}, where $q_a q_b$ specifies the 
flavor content of the interacting $qq$ pair and $\gamma$ the exchanged meson. 
The SU(3)$_F$  is explicitly broken
by the quark masses and by the meson masses. By taking 
$V_{\eta}^{uu}$ = $V_{\eta}^{us}$ = $V_{\eta}^{ss}$
and $V_{\eta'}^{uu}$ = $V_{\eta'}^{us}$ = 0,
one recovers the simpler model of Ref. \cite{CARLSON}
where one does not distinguish between the $uu$, $us$ or $ss$
pairs in the $\eta$-meson exchange.  
Moreover, in Ref. \cite{CARLSON}, for every exchanged meson,  
the radial two-body matrix elements are equal, irrespective of
the angular momentum of the state,  $\ell$ = 0 or $\ell$ = 1.
This is because on takes as parameters the already integrated two-body matrix 
elements of some radial part of the hyperfine interaction, fitted to ground
state baryons. 
Here  one explicitly introduces radial
excitations at the quark level.
\begin{table}
\tbl{Partial contributions from the model Hamiltonian and total energy 
$E =  \sum_{n=1}^5 m_i + \langle T \rangle +
\langle V_{c} \rangle  + \langle V_{\chi} \rangle$ in MeV 
for various $q^4 \overline q$ systems. 
The mass $M$ is obtained from $E$ by 
subtraction of 510 MeV in order to fit the mass of $\Theta^+$. 
The values of the variational 
parameters $\alpha$ and $\beta$
are indicated in the last two columns. }
{\footnotesize
\begin{tabular}{@{}ccccccccc@{}}
\hline
{} &{} &{} &{} &{} &{} &{} &{} &{} \\[-1.5ex]
$q^4 {\overline q} $ &  $\sum_{n=1}^5 m_i$ & $\langle T \rangle $ 
& $\langle V_{c} \rangle $ & $\langle V_{\chi} \rangle$ & $ E $ &
$ M $ & $\alpha(fm)$ & $\beta(fm)$ \\[1ex]
\hline
{} &{} &{} &{} &{} &{} &{} &{} &{} \\[-1.5ex]
${\bf uudd {\overline d}}$ & 1700 & 1864 & 442 & -2044 & 1962 & {\bf 1452} & 0.42 & 0.92 \\[1ex]
${\bf uudd {\overline s}}$ & 1800 & 1848 & 461 & -2059 & 2050 & {\bf 1540} & 0.42 & 1.01 \\[1ex]
${\bf uuds {\overline d}}$ & 1800 & 1535 & 461 & -1563 & 2233 & {\bf 1732} & 0.45 & 0.92\\[1ex]
${\bf uuds {\overline s}}$ & 1900 & 1634 & 440 & -1663 & 2310 & {\bf 1800} & 0.44 & 0.87\\[1ex]
${\bf ddss {\overline u}}$ & 1900 & 1418 & 464 & -1310 & 2472 & {\bf 1962} & 0.46 & 0.92\\[1ex]
${\bf uuss {\overline s}}$ & 2000 & 1410 & 452 & -1310 & 2552 & {\bf 2042} & 0.46 & 0.87\\[1ex]
\hline
\end{tabular}\label{FIVEQ} }
\end{table}
\begin{table}[ph]
\tbl{The antidecuplet mass spectrum (MeV) for  P = + 1.}
{\footnotesize
\begin{tabular}{@{}cccc@{}}
\hline
{} &{} &{} &{} \\[-1.5ex]
${\bf Pentaquark}$ & ${\bf Y,~I,~I_3}$ & ${\bf Present~ results} $ & ${\bf Carlson~ et~ al.}$\\[1ex]
     &   &    Ref. $[\refcite{FS2}]$ & Ref. $[\refcite{CARLSON}]$ \\
\hline
{} &{} &{} &{} \\[-1.5ex]
${\bf \Theta^+}$ & ${\bf 2,0,0}$  & ${\bf 1540}$ & ${\bf 1540}$   \\[1ex]
${\bf N_{\overline {10}}}$        & ${\bf 1,1/2,1/2}$  & ${\bf 1684}$ & ${\bf 1665}$  \\[1ex] 
${\bf \Sigma_{\overline {10}}}$ &  ${\bf 0,1,1}$ & ${\bf 1829}$ & ${\bf 1786}$ \\[1ex]
${\bf \Xi^{--}}$ &  ${\bf -1,3/2,-3/2}$ & ${\bf 1962}$ & ${\bf 1906}$ \\[1ex]
\hline
\end{tabular}\label{ANTIDEC} }
\end{table}
Table \ref{FIVEQ} contains the partial contributions 
and the variational solution $E$ of the 
Hamiltonian \cite{GPP} 
resulting from the trial wave function introduced in Sec. 3.
All specified $q^4 {\overline q}$
systems are needed to construct the antidecuplet and the octet. 
One can see that, except for the
confinement contribution $\langle V_{c} \rangle$, all the other
terms break SU(3)$_F$:  the mass term $\sum_{n=1}^5 m_i$
increases, the kinetic energy $\langle T \rangle$ decreases and the short range 
attraction $\langle V_{\chi} \rangle$ decreases with the quark masses.
For reasons explained in Refs. \cite{FS2,SR} 510 MeV are subtracted from the 
total energy E in order to reproduce the experimental  $\Theta^+$ mass.

For completeness, in the last two columns of Table \ref{FIVEQ}
the values of the variational parameters
$\alpha$ and $\beta$ of the radial wave function
(Sec. 3) are indicated. The parameter $\alpha$  
takes values around $\alpha_0$ = 0.44 fm.
This is precisely the value which minimizes  the
ground state nucleon  mass  when the trial wave 
function is  $\phi \propto \exp[-(x^2 + y^2)/4 \alpha^2_0]$
where $\vec{x}$ and $\vec{y}$ are the 
Jacobi coordinates of Sec. 3.  The quantity $\alpha_0$ 
gives a measure of the 
quark core size of the nucleon because it is its
root-mean-square radius. The parameter  $\beta$  is related to the 
coordinate $\vec{t}$ of the center of mass of $q^4$
relative to $\bar q$. It takes values about twice larger  than  $\alpha$,
which implies that the four quarks cluster together, whereas
$\bar q$ remains separate  in contrast to the diquark Ansatz. 
Table \ref{ANTIDEC} reproduces the antidecuplet mass spectrum 
obtained from the masses M of  Table  \ref{FIVEQ}.
The masses of $\Theta^+$ and $\Xi^{- -}$ can be read off Table \ref{FIVEQ} 
directly. The other masses are obtained from the linear combinations 
\begin{eqnarray}\label{PENTA}
M(N_{\overline {10}}) = 
\frac{1}{3} M(uudd \bar d) + \frac{2}{3} M(uuds \bar s), \nonumber \\
M(\Sigma_{\overline {10}}) = 
\frac{2}{3} M(uuds \bar d) + \frac{1}{3} M(uuss \bar s)~.
\end{eqnarray}
In comparison with  Carlson et al. \cite{CARLSON},
where the mass of $\Theta^+$ is also adjusted to 1540 MeV,
here the masses of $N_{\overline {10}}$,
$\Sigma_{\overline {10}}$ and $\Xi^{- -}$ are higher.
In the lowest order of SU(3)$_F$ breaking, one can parametrize the present 
result by 
the Gell-Mann-Okubo (GMO) mass formula, $M = M_{\overline {10}} + cY$. 
This gives $M \simeq 1829 - 145~ Y$.
The nearly equal spacing between isomultiplets is illustrated in Fig. \ref{fig1} a).
\section{Representation mixing}

The present model contains  SU(3)$_F$ breaking so that representation
mixing appears naturally  and it can be derived dynamically.
Recall that Table 3, column 3 gives the pure antidecuplet masses. 
The pure octet masses are easily calculable using Table \ref{FIVEQ}.
These are
\begin{eqnarray}\label{OCTET}
M(N_{8}) = 
\frac{2}{3} M(uudd \bar d) + \frac{1}{3} M(uuds \bar s) = 
1568 ~\mathrm{MeV}, \nonumber \\
M(\Sigma_8) = 
\frac{1}{3} M(uuds \bar d) + \frac{2}{3} M(uuss \bar s) =
1936 ~\mathrm{MeV}.
\end{eqnarray}
The octet-antidecuplet mixing matrix element $V$ 
has two non-vanishing contributions, one  
coming from the mass term 
and the other from the kinetic energy + hyperfine interaction.  
Its form is 
\begin{equation}\label{COUPLING}
V = \left\{ \renewcommand{\arraystretch}{2}
\begin{array}{cl}
\frac{2 \sqrt{2}}{3} (m_s - m_u) + 
\frac{\sqrt{2}}{3}~[S(uuds \bar s) -  S(uudd \bar d)]  
 = 166  ~\mathrm{MeV}~  
&\hspace{0.5cm} \mbox{for N} \\
\frac{2 \sqrt{2}}{3} (m_s - m_u) +
\frac{\sqrt{2}}{3}~[S(uuss \bar s) -  S(uuds \bar d)] 
= 155  ~\mathrm{MeV}~  
&\hspace{0.5cm} \mbox{for $\Sigma$} 
\end{array} \right. 
\end{equation}
where $S = \langle T \rangle + \langle V_{\chi} \rangle $.
The numerical values on the right hand side of Eq. (\ref{COUPLING})
result from the quark masses $m_{u,d} = 340$ MeV, $m_s = 440 $ MeV
and from the values of $\langle T \rangle $ and $\langle V_{\chi} \rangle $ 
exhibited in Table \ref{FIVEQ}.
One can see that the mass-induced breaking term is identical for
$N$ and $\Sigma$, as expected from simple  
SU(3) considerations, and it
represents more than 1/2 of  V. 
\begin{figure}[ht]
\centerline{\epsfxsize=8cm\epsfbox{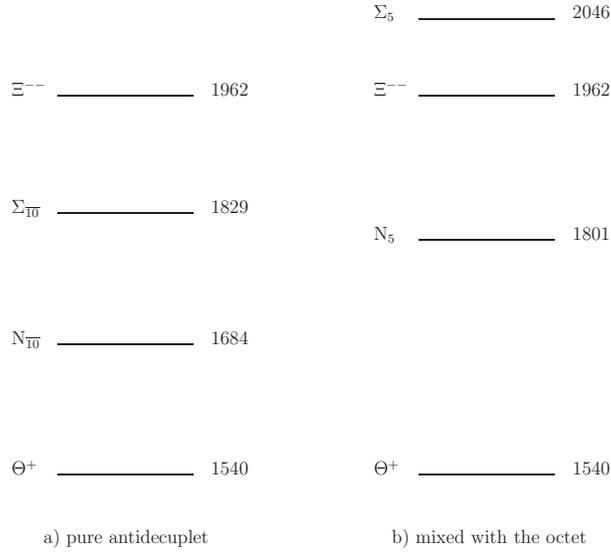}}   
\caption{Comparison between~ a) the pure antidecuplet spectrum of Table \ref{ANTIDEC}
and ~b) the ``mainly antidecuplet'' solutions after the mixing with the octet.
\label{fig1}}
\end{figure}
The masses of the physical states, the ``mainly octet'' $N^*$ and the 
``mainly antidecuplet'' $N_5$, result from diagonalizing 
a 2 $\times$ 2 matrix in each case. Accordingly, the nucleon solutions are
\begin{eqnarray}\label{PHYSN}
N^* =  N_{8} \cos \theta_N - N_{\overline {10}} \sin \theta_N,\nonumber \\
N_5 =  N_{8} \sin \theta_N + N_{\overline {10}} \cos \theta_N,
\end{eqnarray} 
with  the mixing angle defined by
\begin{equation}\label{ANGLE}
\tan 2 \theta_N = \frac{2 V}{M(N_{\overline {10}}) - M(N_{8})}~.
\end{equation}
The masses obtained from this mixing are 1451 MeV and 1801 MeV respectively
and the mixing angle is $\theta_N = 35.34^0$, which means that the
``mainly antidecuplet'' state $N_5$  is  67 \%
 $N_{\overline {10}}$ 
and 33 \%
$N_{8}$, and the ``mainly octet''   $N^*$ state is the other way round.
The latter is located in the Roper resonance mass region
1430 - 1470 MeV.  However this is a 
$q^4 \bar q$ state, i. e.  it is 
different from the $q^3$ radially excited state obtained in Ref. \cite{GPP} 
at 1493 MeV. 
A mixing of the   $q^3$ and 
the $q^4 \bar q$ states could possibly be a better description  of reality. 
The ``mainly antidecuplet'' solution at 1801 MeV is 70 MeV above 
the higher option  of Ref. \cite{ARNDT}, at 1730 MeV, 
interpreted as the  $Y$ = 1 narrow resonance partner of $\Theta^+$.

In a similar way one obtains two $\Sigma$ resonances, the ``mainly octet''
being  at 1719 MeV
and the ``mainly antidecuplet'' at 2046 MeV.
The  octet-antidecuplet mixing angle  is
$\theta_{\Sigma} = - 35.48^0$. The lower state is somewhat above 
the experimental mass range 1630 - 1690 MeV of the 
the $\Sigma(1660)$ resonance. As the higher mass region of $\Sigma$ is
less known experimentally, it would be  difficult to make an assignment for 
the higher state.
The pentaquark spectrum resulting from the octet-antidecuplet mixing is 
illustrated 
in Fig. \ref{fig1} b). One can see that the order of the last
two levels is reversed with respect to case a).

The  mixing angles $\theta_N$ and  $\theta_{\Sigma}$ are nearly equal
in absolute value,  but they have
opposite signs. The reason is that  $M(N_{\overline {10}}) > M(N_8)$
while  $M(\Sigma_{\overline {10}}) < M(\Sigma_8)$. 
Interestingly, each is close to the value of the ideal mixing angle 
$\theta_N^{id}$ = 35.26$ ^0$ and $\theta_{\Sigma}^{id} = - 35.26^0$.
This implies that in practice the ``mainly antidecuplet'' $N_5$ state carries 
the whole hidden strangeness and that $N^*$ has a simple  content,
for example $uudd \bar d$ when the charge is positive.

\section{Heavy pentaquarks}
\begin{table}[h]
\label{HEAVY}
\tbl{Masses (MeV)~ of~ the~ positive parity antisextet charmed pentaquarks.}
{\begin{tabular}{@{}ccccc@{}}\toprule
${\bf Pentaquark}$ & ${\bf I}$ & ${\bf Content}$ & ${\bf FS~ model}$   & ${\bf Lattice}$ \\
      &         &                           &  Ref.~$[\refcite{FS1}]$ & Ref.~$[\refcite{CH2}]$ \\
\hline
{} &{} &{} &{} &{} \\[-1.5ex]
${\bf \Theta^0_c}$ & ${\bf 0}$    & ${\bf u~ u~ d~ d~ \bar c}$ & ${\bf 2902}$  & ${\bf 2977\pm104}$ \\[1ex]
${\bf N^0_c}$      & ${\bf 1/2}$  & ${\bf u~ u~d~s~\bar c}$    & ${\bf 3161}$  &${\bf 3180\pm69}$  \\[1ex] 
${\bf \Xi^0_c}$    &  ${\bf 1}$   & ${\bf u~ u~ s~ s~ \bar c}$ & ${\bf 3403}$  &${\bf 3650\pm95}$ \\
{} &{} &{} &{} &{} \\[-1.5ex]
\hline
\end{tabular}}
\end{table}
\noindent
Based on the same constituent quark model \cite{GPP}, 
positive parity heavy charmed pentaquarks of minimal content $uudd \bar c$ 
have been proposed \cite{FS1} long before the first 
observation \cite{NAKANO} of $\Theta^+(uudd \bar s)$. Table 4
reproduces the results of Ref. \cite{FS1} where the masses represent the binding
energies $\Delta E$ (Table II) to which threshold energies $E_{T}$,
(Table I) have been added. These results are compared 
with the only
lattice calculations which predict positive parity \cite{CH2}.
Interestingly the masses are quite similar in the two approaches. 
In the FS model \cite{GPP} the lightest negative parity pentaquark is
a few hundreds MeV heavier than $\Theta^0_c$ of Table 4. 
The experimental search 
for charmed pentaquarks is contradictory so far \cite{WANG}.

\section{Conclusions}

In the new light shed by the pentaquark studies, 
the usual practice of hadron spectroscopy is expected to change.
There are hints that the wave functions 
of some excited states might contain $q^4 \overline{q}$ components.
These components, if obtained quantitatively, would perhaps better explain the 
widths and mass shifts in the baryon resonances. In particular
the mass of the Roper resonance may be further shifted up or down.
Also it is important to understand the role of the chiral symmetry breaking 
on the properties of pentaquarks \cite{HOSAKA}, inasmuch as 
the predictions of Ref. \cite{DPP},
which motivated this new wave of interest,
are essentially based on this concept.      

\end{document}